\documentclass[twoside,slac_one]{revtex4}
\usepackage{epsfig}
\usepackage{graphicx}
\usepackage{fancyhdr}
\usepackage{bm}
\usepackage{latexsym}
\usepackage{lscape}

\newcommand{\be}{\begin{equation}}
\newcommand{\ee}{\end{equation}}
\newcommand{\bea}{\begin{eqnarray}}
\newcommand{\eea}{\end{eqnarray}}
\newcommand{\Tr}{{\rm Tr}}

\begin{document}

\title{Recent results on large N gauge theories on a single site
  lattice with adjoint fermions}

%

\author{R. Narayanan}
\affiliation{Department of Physics, Florida International University,
 Miami, FL 33199, USA.}

\begin{abstract}
Large N gauge theories with adjoint matter can be numerically studied
using lattice techniques. Eguchi-Kawai reductions holds for this
theory and one can reduce the lattice model to a single site. Hybrid
Monte Carlo algorithm can be used to simulate this model. One can
either perform an exact computation of the ``fermionic force'' or
use pseudo fermions as part of the HMC algorithm. The former algorithm
is slower than the latter but has the advantage that one can work with
any real number for the fermion flavor. Some results using both
algorithms
will be presented.
\end{abstract}

\maketitle

\thispagestyle{fancy}


\section{Introduction}
Lattice studies of
vector like gauge theories with adjoint fermion matter 
with the aim of understanding the conformal window has recently
attracted considerable attention (see~\cite{DelDebbio:2010zz} and
references therein). 
The gauge group is chosen to be SU(N) and
the beta function is
\be
\frac{d \alpha}{d \ln a^2} = \beta(\alpha) = \frac{11-4f}{3} \alpha^2 +
\frac{34-32f}{3}\alpha^3 + \cdots;\ \ \alpha=\frac{\lambda}{16\pi^2};
\ee
where $a$ is the lattice spacing, $f$ is the number of Dirac flavors
or adjoint fermions and $\lambda=g^2N$ is the 't Hooft gauge
coupling 
on the lattice. The first two coefficients in the beta function are
renormalization scheme independent.
In order to maintain asymptotic freedom, we restrict ourselves to $f<3$.
The two loop beta function
has a zero if $f=2$ and this has a motivated numerical studies of
SU(2) gauge group with two Dirac flavors of fermions in the
  adjoint
  representation~\cite{Catterall:2011zf,DeGrand:2011qd,Hietanen:2009az}.

A continuum analysis of the theory with adjoint fermions
on ${\bf R}^3\times S^1$
with periodic boundary conditions for fermions
in the compact direction shows that the $Z_N$ symmetry
is not broken in that direction~\cite{Kovtun:2007py}. 
An analysis on $S^3\times S^1$ also shows a region where
the $Z_N$ symmetry is not broken~\cite{Hollowood:2009sy} .
A lattice
analysis of the same theory with Wilson fermions indicates
that one can reduce the compact direction to a single site
on the lattice and still maintain the $Z_N$ symmetry
\cite{Bringoltz:2009mi,Bringoltz:2009fj}.
This is expected to be the case for $f\ge \frac{1}{2}$~\cite{Hietanen:2009ex} and
for
non-zero quark masses \cite{Azeyanagi:2010ne,Hietanen:2010fx,Catterall:2010gx}.

The model on the single site lattice is
defined in Sec.~\ref{model}. We will use one-loop perturbation theory
of this model to show that it is expected to the correct continuum
phase in Sec.~\ref{oneloop}.
A summary of our non-perturbative results will be presented in Sec.~\ref{results}.

\section{The model}\label{model}
The action on a single site lattice with
one flavor of adjoint Dirac overlap fermion is given by
\be
S = S_g + S_f.
\ee
The matrices, $H_\mu$; $\mu=1,2,3,4$
are elements of the $su(N)$ algebra and conjugate to the four SU(N)
gauge degrees of freedom, $U_\mu$; $\mu=1,2,3,4$.
The gauge action is 
\be
S_g = -bN \sum_{\mu\ne\nu=1}^4
\Tr  \left [ U_\mu U_\nu U_\mu^\dagger U_\nu^\dagger\right].
\label{gaction}
\ee
The lattice gauge coupling constant is $b=\frac{1}{g^2N}$.
\footnote{This coupling is related to the standard lattice coupling,
  $\beta$, by $b=\frac{\beta}{2N^2}$. A value of $b=0.35$ corresponds
to $\beta=2.8$ for $N=2$ and $\beta=6.3$ for $N=3$.}
The overlap fermion action is
\be
S_f =  -2f\log\det H_{o+}.
\label{oaction}
\ee 

The Hermitian 
massive overlap Dirac operator is defined by \cite{Edwards:1998wx,Neuberger:1997fp}
\be
H_o(\mu)= \frac{1}{2}\left [ \left( 1 + \mu \right)\gamma_5 +
\left(1-\mu\right)\epsilon(H)\right],\label{hover}
\ee
where $\mu\in[0,1]$ is the bare mass
and
\be
H_{o\pm}^2(\mu) = \frac{1+\mu^2}{2}P_\pm \pm \frac{1-\mu^2}{2} P_\pm
\epsilon(H) P_\pm; \ \ \  P_\pm=\frac{1\pm\gamma_5}{2},
\label{hoversq}
\ee
factorizes into two disjoint pieces corresponding to the two
chiralities. 
Note that $f=1$ in (\ref{oaction}) is the correct result for a single Dirac fermion in the
adjoint representation~\footnote{We are assuming that global topology
  is completely suppressed and one can restrict the theory to the zero
  topological sector.}. We can set $f$ to be half integers and
simulate Majorana fermions but we can also extend $f$ to any real number
in (\ref{oaction}).

The function $\epsilon(H)$ appearing in (\ref{hover}) is the sign function of the Hermitian
Wilson Dirac operator, $H$.
The Hermitian Wilson Dirac operator 
for adjoint fermions is
given by
\bea
H &=& \pmatrix{ 4 - m -\frac{1}{2}\sum_\mu \left( V_\mu + V_\mu^t\right)
& \frac{1}{2}\sum_\mu \sigma_\mu \left(V_\mu - V_\mu^t\right) \cr
-\frac{1}{2}\sum_\mu \sigma^\dagger_\mu \left(V_\mu - V_\mu^t\right) &
-4 + m +\frac{1}{2}\sum_\mu \left( V_\mu + V_\mu^t\right)\cr}\cr
&=& (4-m)\gamma_5
- \sum_\mu \left ( w_\mu V_\mu + w_\mu^\dagger V_\mu^t\right)
\label{wilson}
\eea
with 
\be
w_\mu = \frac{1}{2}
\pmatrix { 1 & -\sigma_\mu \cr \sigma_\mu^\dagger & -1\cr}.
\ee
Let $\Phi$ be a traceless Hermitian matrix and denote one component of
an adjoint Dirac fermion on the single site lattice.
The action of $V_\mu$ on $\Phi$ is given by
\be
V_\mu \Phi = U_\mu \Phi U_\mu^\dagger;\ \ \ 
V^t_\mu \Phi = U^\dagger_\mu \Phi U_\mu.\label{vaction}
\ee

One can verify that $H$ is Hermitian in the usual sense:
\be
\Tr \Psi^\dagger H \Phi = \left[ \Tr \Phi^\dagger H \Psi \right]^* =
\Tr \left[ (H\Psi)^\dagger \Phi\right].
\ee
Therefore $\Psi^\dagger H = (H\Psi)^\dagger$ and
in addition it is also true that $\Tr H\Phi=0$ if $\Tr\Phi=0$.
The same is also true for $H_o(\mu)$.

\section{Weak coupling perturbation theory}\label{oneloop}

For the single site lattice theory to reproduce the correct infinite
volume continuum theory, the center symmetry that takes
\be
U_\mu \to e^{i\frac{2\pi k_\mu}{N}} U_\mu;\ \ \ \  k_\mu=0,\cdots,N-1
\ee
with $k_\mu$; $\mu=1,2,3,4$ independent of each other should not be
broken.
In the limit of large $N$, this amounts to saying that $\Tr U_\mu =0$
for
all $\mu$ which is equivalent to the statement that the eigenvalues of
$U_\mu$ are uniformly distributed on the unit circle. The single
site perturbation theory is given by
\be U_\mu = e^{ia_\mu} D_\mu e^{-ia_\mu};\ \ \ \ 
D_\mu^{ij}=e^{i\theta_\mu^i}\delta_{ij}. 
\ee

Keeping $\theta_\mu^i$ fixed, we expand in powers of
$a_\mu$.
The lowest contribution to $S_g$ comes from the quadratic term in
$a_\mu$~\cite{Bhanot:1982sh} and the lowest contribution
to $S_f$ comes from setting $a_\mu=0$.
Each $V_\mu$ has $\frac{N(N-1)}{2}$
two by two blocks of the form
\be
\pmatrix{
\cos(\theta_\mu^i-\theta_\mu^j) &   
\sin(\theta_\mu^i-\theta_\mu^j)   \cr
-\sin(\theta_\mu^i-\theta_\mu^j)   &
\cos(\theta_\mu^i-\theta_\mu^j)   \cr}
\ee 
with $1\le i < j \le N$. The remaining $(N-1)\times (N-1)$ matrix is a
unit matrix.
Therefore, the gauge field effectively has $(N-1)$ zero momentum modes
and $N(N-1)$ non-zero momentum modes of the form
$e^{i(\theta_\mu^i-\theta_\mu^j)}$ with $1\le i \ne j \le N$.
If $\theta_\mu^i$ for a fixed $\mu$ are uniformly distributed on the
unit circle and there is no correlation between the different $\mu$,
the single site model will correctly reproduce the momentum integral
of the
infinite volume continuum theory. Our aim in one-loop perturbation
theory is to study the distribution of $\theta_\mu^i$.

The computation of the fermion determinant
reduces to a free field calculation at this order
and the result is
\be 
S_f = -4f \sum_{i\ne j} \ln \lambda( \theta^i-\theta^j+\phi)-4(N-1)f\ln\lambda(\phi)
\ee 
where $e^{i\phi_\mu}$,
$\phi_\mu = \frac{2\pi k_\mu}{N}$, is the phase associated with the boundary
condition in the $\mu$ direction.
The eigenvalues, $\pm\lambda(p)$, are two fold degenerate and
given by
\be
\lambda(p) =  \sqrt{\frac{1+\mu^2}{2} + \frac{1-\mu^2}{2} 
\frac{2\sum_\mu\sin^2\frac{p_\mu}{2} -m}
{\sqrt{
\left(2\sum_\mu\sin^2\frac{p_\mu}{2} -m\right)^2 + \sum_\mu \sin^2 p_\mu}}}.\label{lamoverlap}
\ee

The complete result from fermions and gauge fields is
\be
S =  \sum_{i\ne j} \left\{
\ln \left[\sum_\mu \sin^2 \frac{1}{2}\left(\theta_\mu^i-\theta_\mu^j\right) \right]
- 4f
\ln \lambda( \theta^i-\theta^j+\phi)\right\} -4(N-1)f\ln\lambda(\phi).
\label{pertact}
\ee
If $f=0$, the minimum of the action occurs when all $\theta_\mu^i=0$
and the single site model is not in the correct continuum phase
~\cite{Bhanot:1982sh}.
If $\mu=0$, $S_f\to\infty$ when all $\theta_\mu^i=0$ and this choice
need not be the minimum.

Overlap fermions reproduce the correct continuum behavior by
restricting the
full Brillouin zone to a physical region around zero defined by
\be
m > 2\sum_\mu\sin^2\frac{p_\mu}{2} = 2\sum_\mu\sin^2\frac{\theta_\mu^i-\theta_\mu^j}{2}.
\ee
We cannot set $m$ to be very large since overlap fermions reduce to
na\"ive fermions as $m\to\infty$ and na\"ive fermions on a single site
lattice do not reproduce the correct continuum
behavior~\cite{Hietanen:2009ex}.
We cannot set $m$ to be too small since we will not cover a
substantial region of the Brillouin zone to realize the correct
momentum measure. 

Unfortunately (\ref{pertact}) is a function of $4N$
variables and it is not easy to study it analytically and find the
minimum. One option is to numerically study this function.
In order to find the minimum of $S$, we consider the Hamiltonian
\be
H = \frac{1}{2}\sum_{\mu,i} \left(\pi_\mu^i\right)^2 + \beta S.
\ee
For large $\beta$, the Boltzmann measure $e^{-H}$ will be dominated
by the minimum of $S$. We can perform a HMC update of the $\pi,\theta$
system to find this minimum~\cite{Hietanen:2009ex}. 

\begin{figure}[ht]
\centering
\includegraphics[width=80mm]{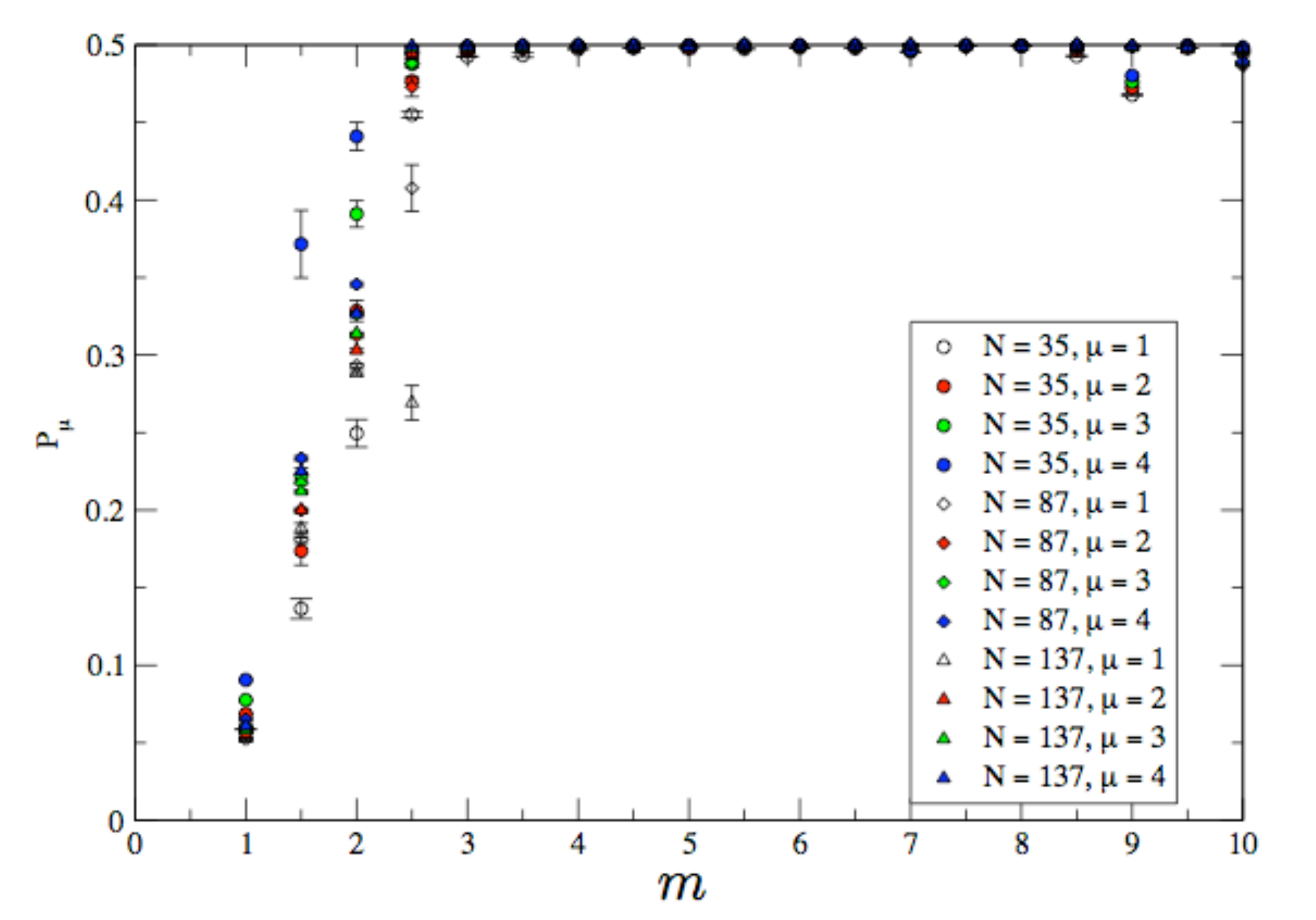}
\includegraphics[width=80mm]{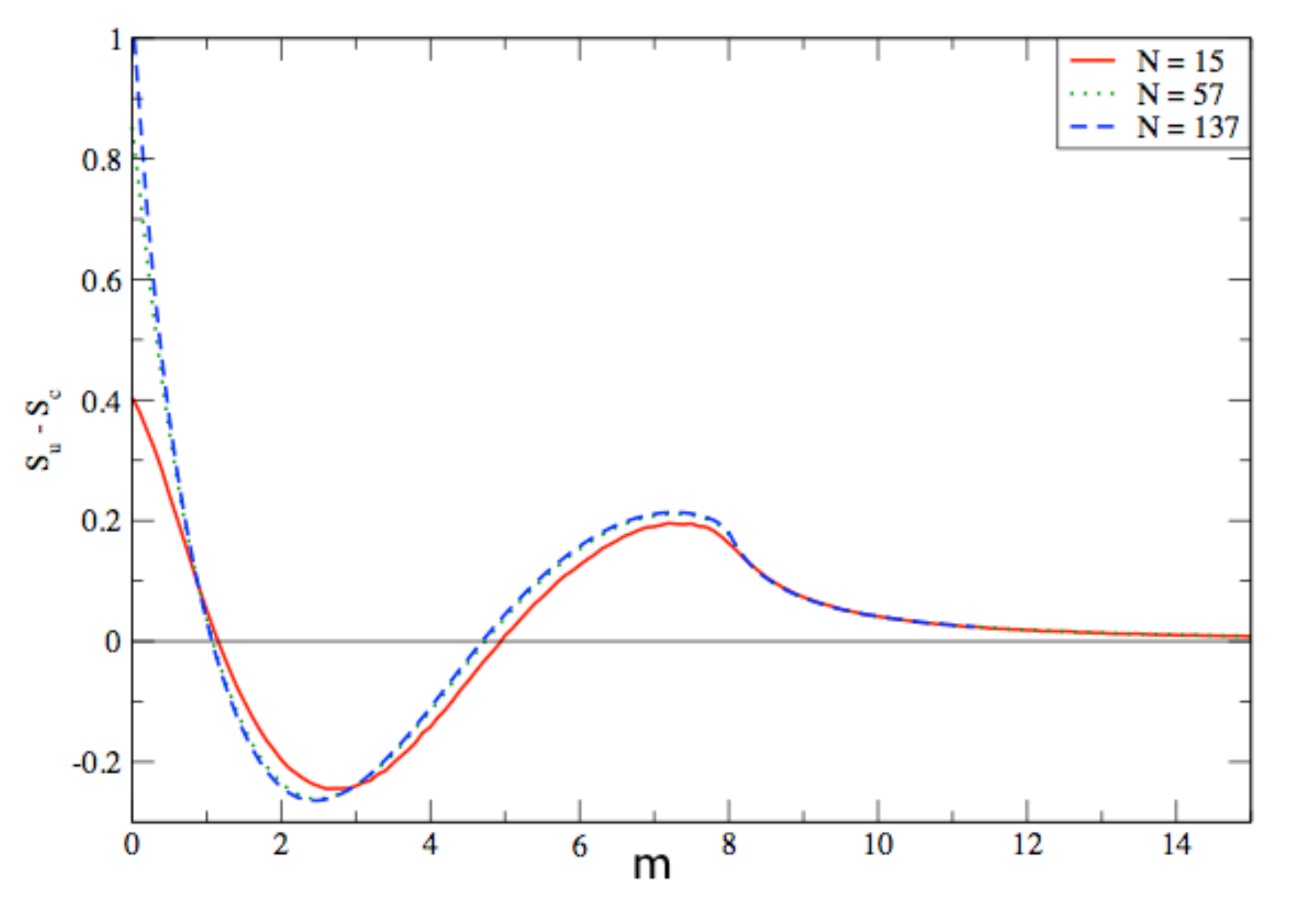}
\caption{ The left panel is a 
plot of $P_\mu$ as a function of $m$ for several different values of
$N$.
The right panel shows a plot of $S_u-S_c$ as a function of $m$ for several different values of $N$.
} \label{fig1}
\end{figure}

A choice for
the order parameters associated with the $Z_N^4$ symmetries is~\cite{Bhanot:1982sh}
\be
P_\mu = \frac{1}{2} \left( 1 - \frac{1}{N^2}|\Tr U_\mu|^2\right)
=\frac{1}{N^2}\sum_{i,j}
\sin^2 \frac{1}{2}\left(\theta_\mu^i-
\theta_\mu^j\right) 
\ee
If $P_\mu=\frac{1}{2}$, then the $Z_N$ symmetry in that direction is
not broken. 
If $\theta_\mu^i$ are uniformly distributed in a width $\alpha\le
2\pi$, then
\be
\lim_{N\to\infty} P_\mu= \frac{1}{2}\left[ 1-
 \left(\frac{2}{\alpha}\sin\frac{\alpha}{2}\right)^2\right].\label{palpha}
\ee
The result of the numerical simulation to look for the minimum of $S$
in (\ref{pertact}) with $f=1$
is shown in the left panel of Fig.~\ref{fig1}. It is clear that we will not reproduce the
correct continnum behavior if $m<3$. It seems that one can take $m$ as
large as $8$. This analysis does not take into account possibilities
of
correlations in the different directions. Instead of studying
correlations
in different directions, we 
compute the correlated action, $S_c$, with 
$\theta^j_\mu=\frac{2\pi j}{N}$ and compare it to
the uncorrelated action, $S_u$, with
$\theta^j_\mu=\frac{2\pi \pi^\mu_j}{N}$ where $\pi^\mu$ are different
permutations for different $\mu$. The difference between $S_u$ and
$S_c$ is plotted as a function of $m$ for several values of $N$ in the
right panel of
Fig.~\ref{fig2} for $f=1$. It shows that the correlated one is below the
uncorrelated one for $m>5$. Using the above two arguments, we conclude
that we need to set $3 < m < 5$ in order for the single site theory to
reproduce the correct
momentum
measure when $f=1$.

\section{Non-perturbative results}\label{results}
We need to verify if the results obtained in one-loop perturbation
theory in the previous section remains valid in a full
non-perturbative computation. The non-perturbative computation
is performed using the Hybrid Monte Carlo algorithm. 
Let $H_\mu$ be traceless Hermitian matrices that are conjugate to
the gauge fields, $U_\mu$.
The algorithm starts with one choice for $U_\mu$. Then, we draw
$H_\mu$ according to a Gaussian distribution. 

The equations of motion for $U_\mu$ are
\be
\frac{ d U_\mu}{d\tau} = i H_\mu U_\mu.\label{ueqn}
\ee
Setting $\frac{d S}{d\tau}=0$ 
results in
\be
\sum_{\mu=1}^4 \Tr \left [ H_\mu \frac{ dH_\mu}{d\tau} \right]+
\frac{ dS_g}{d\tau} + \frac{ dS_f}{d\tau}=0.
\ee
The derivative of $S_f$ with respect to $\tau$ is referred to as
fermionic force term and is computationally intensive. One can derive
exact expressions for the single site model~\cite{Hietanen:2009ex}.
It involves exact diagonalization of $H$ and the computational cost
grows like $N^6$. The advantage of computing the fermionic force
exactly is that one can work with any real value of $f$.
An alternative approach is to use the pseudo-fermion algorithm to
compute the fermionic force term~\cite{inprep}. This is less
computationally
intensive but works only for integer values of $f$.

\begin{figure}[ht]
\centering
\includegraphics[width=80mm]{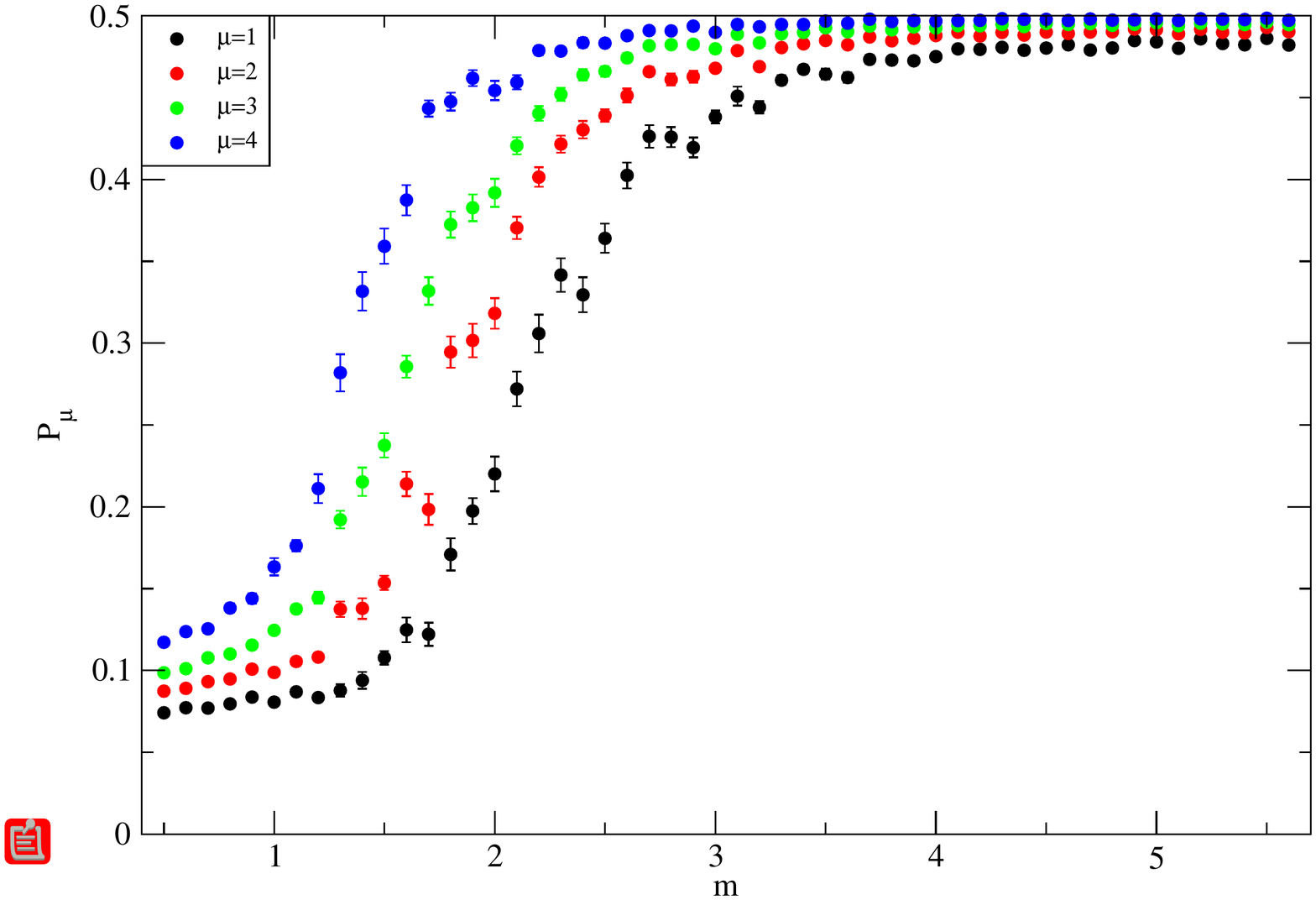}
\includegraphics[width=80mm]{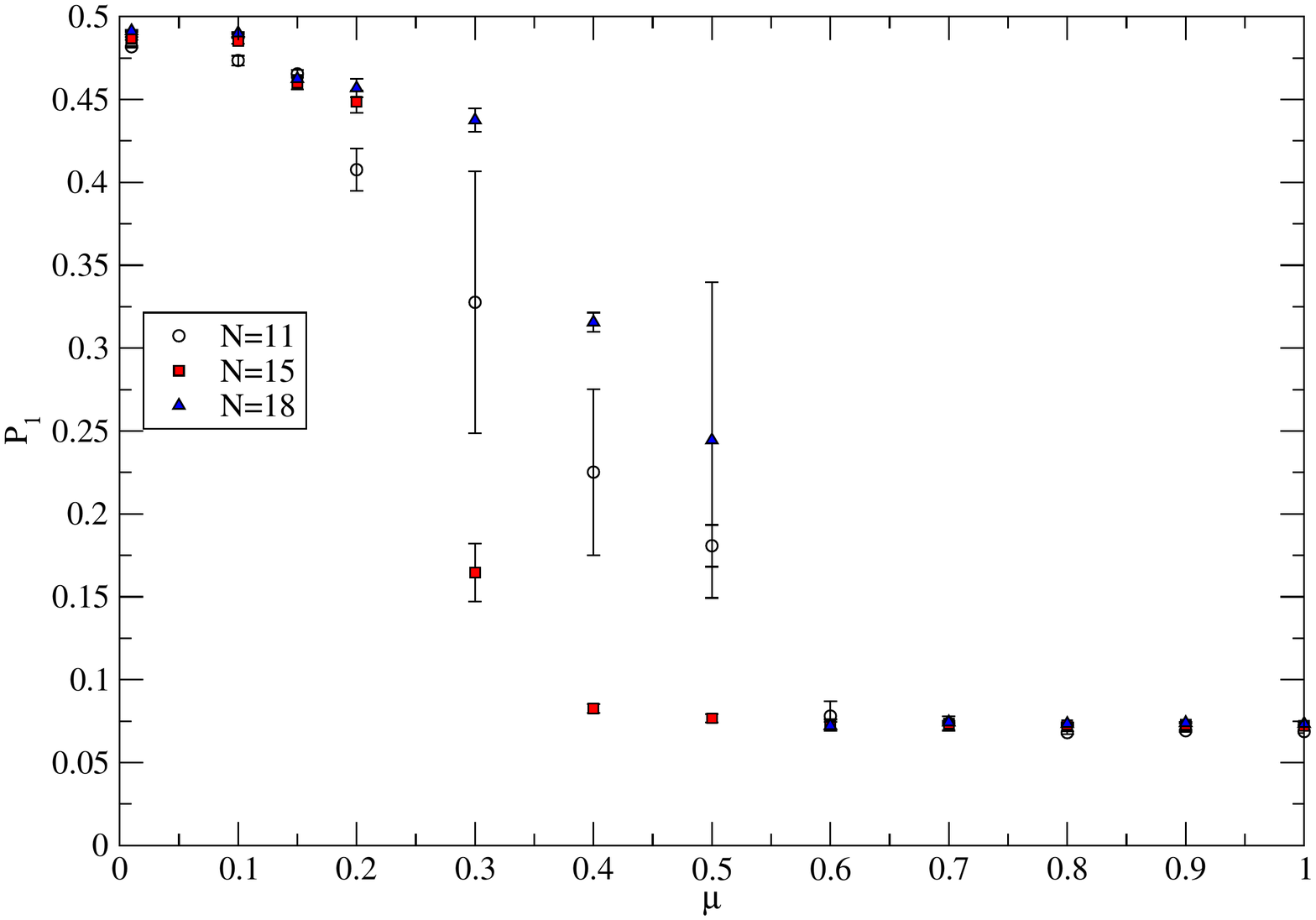}
\caption{ The left panel is a 
plot of $P_\mu$ as a function of $m$ in the full numerical simulation
with $N=11$, $b=7$, $f=\frac{1}{2}$ and $\mu=0.01$.
The right panel shows a plot of $P_1$ (this is the one that is broken
first)
as a function of $\mu$ for several different values of $N$.
} \label{fig2}
\end{figure}

The result for massless fermions (we set $\mu=0.01$ in the numerical
simulation)
is shown in the left panel of Fig.~\ref{fig2} using the exact
algorithm for the fermionic force. We can see
that the theory will reproduce the correct continuum limit if
$m>3$. We set $m=5$ and studied the behavior of the model as a
function
of the quark mass. The result is plotted in the right panel of
Fig.~\ref{fig2}.
We see that the model will reproduce correct continuum physics even
when the fermions are massive. 

\subsection{Chiral symmetry}

In order to study whether chiral symmetry is spontaneously broken, we 
studied the behavior of the low lying positive eigenvalues, $0<
\lambda_1 < \lambda_2 \cdots$, of the
hermitian
overlap Dirac operator. 
If chiral symmetry is broken, we expect a relation of the form
\be
z_i = N^2 \lambda_i \Sigma(b)
\ee
where
the joint distribution of the scaled variables, $z_i$, are given by
some chiral random matrix model~\cite{Verbaarschot:2000dy}
and $\Sigma(b)$ is the value of the chiral condensate. 

\begin{figure}[ht]
\centering
\includegraphics[width=80mm]{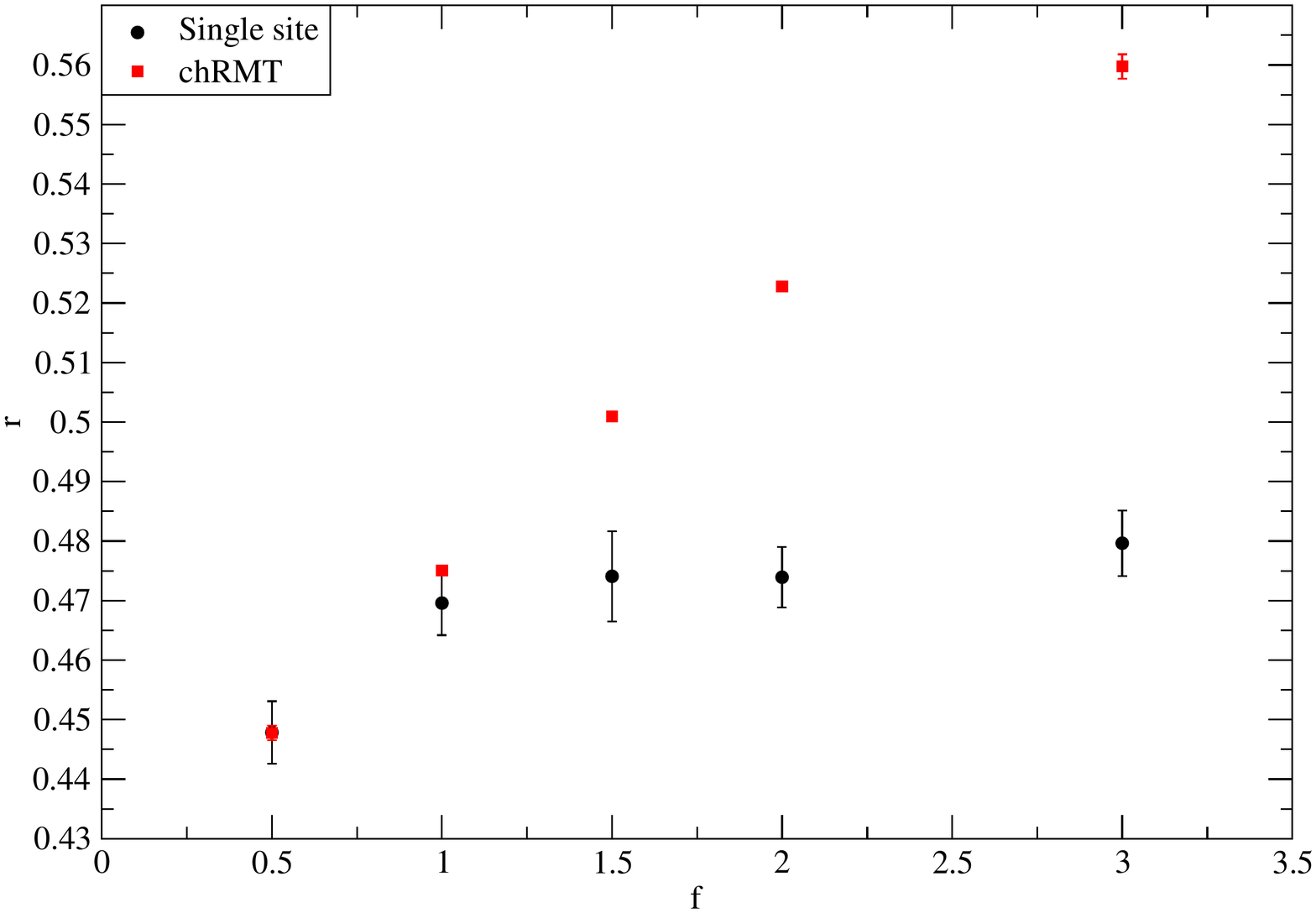}
\includegraphics[width=80mm]{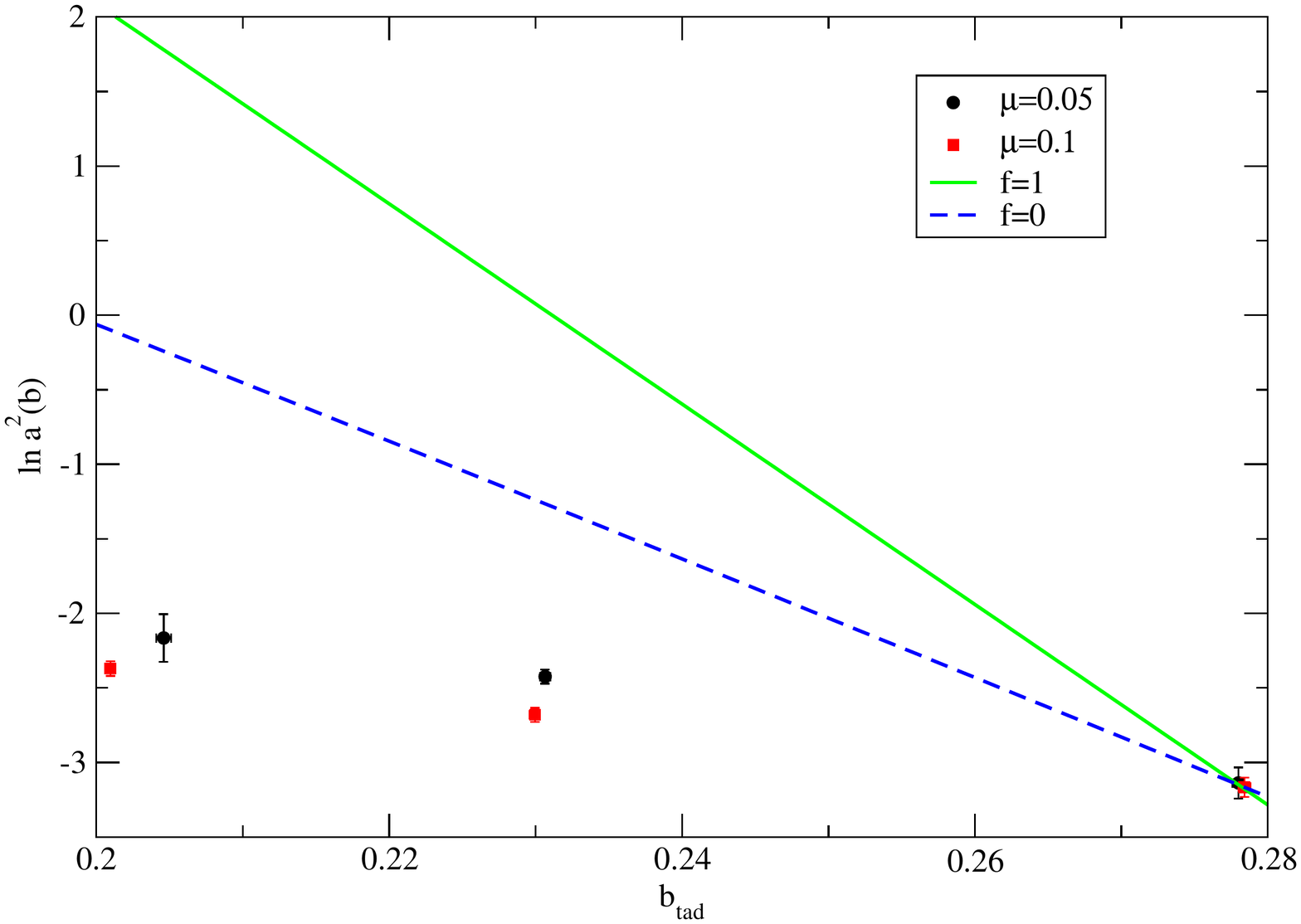}
\caption{ The left panel is a 
plot of $r$ as a function of $f$ in the full numerical simulation
with $N=11$, $b=5$, and $\mu=0.01$.
The right panel shows the lattice scale as a function of the lattice
coupling
for $f=1$. The two loop result for $f=0$ and $f=1$ are shown for
comparison.
} \label{fig3}
\end{figure}

The chiral Random Matrix theory ensemble for a symplectic 
matrix, $C=\sum_\mu \sigma_\mu C_\mu$, is
\be 
Z = \int [dC_\mu] e^{-\sum_\mu \sum_{ij} 
\left[ C_\mu^{ij}\right]^2}
\left[ \det H_{\rm rmt}\right]^f;\ \ \ \ 
H_{\rm rmt}=\pmatrix{ \mu & C \cr C^\dagger & -\mu \cr}
\ee 
with $C_\mu$ being a real square matrix. 
We expect $z_i$ to be 
the eigenvalues of $H_{\rm rmt}$ with $\Sigma(b)$ being the scale
that relates these eigenvalues to the eigenvalues of $H_o$.
We want to eliminate the scale set by the chiral condensate, 
and we focus on 
\be 
r= \left\langle \frac{\lambda_1}{\lambda_2} \right\rangle.
\ee 
The result as a function of $f$ with $b=5$, $m=5$, $\mu=0.01$ and
$N=11$ 
is shown in the left panel of Fig.\ref{fig3}. By comparion with
chiral random matrix theory, it looks like chiral symmetry is broken
for $f=0$ and $f=1$ but not for $f>1$. This would be the case if the
non-pertubative
beta function has a zero for $f>1$.

\subsection{Setting the scale}

In order to get some insight into the non-perturbative beta function,
we study a lattice scale as a function of the lattice coupling.
An $L\times T$ Wilson loop operator in the $\mu-\nu$ plane is given by
\be
W(L,T) = U_\mu^L U_\nu^T {U_\mu^\dagger}^L {U_\mu^\dagger}^T.
\ee
The eigenvalues, $e^{i\theta_k}$; $k=1,\cdots,N$ of this operator are
gauge invariant. Let $p(\theta;L,T,b)$ be the distribution of these
eigenvalues
with $\theta\in[-\pi,\pi)$.
This distribution undergoes a transition~\cite{Narayanan:2006rf} at $N\to\infty$ as the area,
$LT$, is changed at a fixed coupling $b$: the
distribution has a gap at $\pi$ for small areas and it becomes gapless
for large areas. There is a critical area $A_c(b)$ where the gap
closes.
There is a
universal
function~\cite{Narayanan:2007dv} describing the distribution in terms of the scaled variables
derived from $A(b)$ and $\theta$ in the vicinity of $A_c(b)$ and
$\pi$. 

Let
\be
O_N(y;A,b) = \langle \det \left( e^{\frac{y}{2}} + e^{-\frac{y}{2}}
    W(L,T)\right)\rangle;\ \ \ \ A=LT.
\ee
The region close to $y=0$ probes $\theta$ close to $\pi$.
Let 
\be
O_N(y;A,b) = C_0(A,b,N) + C_1(A,b,N) y^2 + C_2(A,b,N) y^4 + \cdots.
\ee
It is useful to define
\be
\Omega(A,b,N) = \frac{ C_0(A,b,N) C_2(A,b,N}{C_1^2(A,b,N)}.
\ee
One can show using the universal scaling function that
\be
\Omega(A_c(b),n,\infty) =
\frac{\Gamma^4\left(\frac{1}{4}\right)}{48\pi^2} = 0.364739936
\ee
We can define $A_c(b,N)$ at a fixed $N$ and $b$ as the area where 
\be
\Omega(A_c(b,N),b,N)=0.364739936,\label{tranval}
\ee
and
\be
\lim_{N\to\infty} A_c(b,N) = A_c(b),
\ee
will be the location of the transition at infinite $N$.

Since we are working at a fixed but large $N$ in this paper, we will
define our length scale as
\be
a(b) = \frac{1}{\sqrt{A_c(b,N)}}.
\ee

It is necessary to work with a lattice coupling that shows the weak to
strong coupling transition that is essentially free of finite $N$ effects.
We chose $N=18$ and set the lattice couplings to $b=0.32, 0.35, 0.4$. The fermion
mass was set to $\mu=0.1$ and
$\mu=0.05$ and we set $m=4$. The pseudo-fermion algorithm was used to
compute the fermionic force.

This is an irrelevant parameter but needs to be chosen in a specific
range
to realize the correct continuum limit. Based on previous studies~\cite{Hietanen:2010fx}, we
set $m=4$ in this paper.

We restrict ourselves to square Wilson
loops and this enables us to extract a length scale
using
linear interpolation. The results are plotted in the right panel of
Fig.~\ref{fig3}
and compared with the two loop result for $f=0$ and $f=1$. The length
scale
changes very little in the range of coupling we have studied in this
paper.
Let us assume a beta function of the form
\be
\beta(\alpha) = \epsilon + (\alpha-\alpha_0)^2
\ee
motivated by~\cite{Kaplan:2009kr}. Assume $\epsilon>0$ but small for our case
since the two loop beta function has a zero if $f$ is slightly bigger
than unity. The scale as a function of the coupling is given by
\be
\ln a^2 =
\frac{1}{\sqrt{\epsilon}}\tan^{-1}\frac{\alpha-\alpha_0}{\sqrt{\epsilon}}.
\ee
If $\epsilon$ is small and our lattice coupling is larger than
$\alpha_0$ and not close to it, the scale will change very little if we change the
coupling.
This leads us to speculate that the single site model we are simulating might be close
to a situation where the beta function has a zero. A careful analysis
of the large $N$ corrections along with results at weaker coupling are
needed to confirm this speculation. In addition, it will be necessary
to study the chiral limit. The effect due to fermion masses in the
right panel of Fig.~\ref{fig3} are small.

\begin{acknowledgments}
R.N. acknowledges partial support by the NSF under grant number
PHY-0854744.  R.N. would like to acknowledge ongoing collaboration
with
Ari Hietanen.
\end{acknowledgments}

\bigskip 

\end{document}